\documentclass[letterpaper, 10 pt, conference]{ieeeconf}  

\IEEEoverridecommandlockouts                              

\usepackage{comment}
\usepackage{cite}
\usepackage{soul}
\usepackage{xcolor}
\usepackage{amsmath}
\usepackage{amssymb}
\usepackage{subcaption}
\usepackage[]{draftfigure}
\setkeys{Gin}{width=\linewidth}
\usepackage{multirow}
\usepackage{graphicx}
\usepackage{float}
\usepackage{mathtools}

\usepackage{bm}
\makeatletter
\let\NAT@parse\undefined
\makeatother
\usepackage{hyperref}
\hypersetup{
   colorlinks=true,
    linkcolor= blue,
    allcolors=blue,
    citecolor = blue,
    filecolor=black,      
    urlcolor=blue,
    }    
    
\usepackage{todonotes}

\usepackage [autostyle, english = american]{csquotes}
\MakeOuterQuote{"}

\usepackage{lineno}

\newcommand{\RNum}[1]{\uppercase\expandafter{\romannumeral #1\relax}}
\newcommand{\eat}[1]{}
\newcommand{\plus}{\raisebox{.4\height}{\scalebox{.6}{+}}}

\DeclareMathOperator*{\argmaxA}{arg\,max} 

\title{\LARGE \bf
A Multi-Agent Deep Reinforcement Learning Coordination Framework for Connected and Automated Vehicles at Merging Roadways}

\author{Sai Krishna Sumanth Nakka, \emph{IEEE Student Member}, Behdad Chalaki, \emph{IEEE Student Member},\\ Andreas A. Malikopoulos, \emph{IEEE Senior Member}
\thanks{This research was supported by ARPAE's NEXTCAR program under the award number DE-AR0000796.}
\thanks{The authors are with the Department of Mechanical Engineering, University of Delaware, Newark, DE 19716 USA (emails: \texttt{\{nakkash;bchalaki;andreas\}@udel.edu}).}}%

\begin{document}

\maketitle
\thispagestyle{empty}
\pagestyle{empty}

\begin{abstract}

The steady increase in the number of vehicles operating on the highways continues to exacerbate congestion, accidents, energy consumption, and greenhouse gas emissions. Emerging mobility systems, e.g., connected and automated vehicles (CAVs), have the potential to directly address these issues and improve transportation network efficiency and safety. In this paper, we consider a highway merging scenario and propose a framework for coordinating CAVs such that stop-and-go driving is eliminated. We use a decentralized form of the actor-critic approach to deep reinforcement learning\textemdash multi-agent deep deterministic policy gradient. We demonstrate the coordination of CAVs through numerical simulations and show that a smooth traffic flow is achieved by eliminating stop-and-go driving.

\end{abstract}

\section{INTRODUCTION}
\PARstart{T}{he} disproportionate growth in traffic volume compared to the road capacity causes an increase in congestion on highways with significant implications on the road safety and environmental footprint of road vehicles \cite{Schrank2019}. Traffic bottlenecks usually cause congestion; for instance, recurrent bottlenecks such as merging on highways are responsible for around $40$-$80$\% of congestion in the US \cite{Spiller2017}. In terms of safety, $1.7$\% of the total number of fatal crashes are caused by lane change/merging maneuvers \cite{safetyfacts2017}.
Data collected by NHTSA \cite{crashstats2015} shows that a major portion of the accidents is caused due to human error. 
By combining communication technologies with the capabilities of automated vehicles, we can improve both safety and efficiency \cite{zhao2019enhanced}.
Connected and automated vehicles (CAVs) are able to interact with each other to gather information and make decisions to reduce the potential conflicts by $90-94$\% \cite{papadoulis2019evaluating}. In addition to increasing road safety, CAVs can enable significant improvements in traffic efficiency, energy consumption, and ultimately reduce the carbon footprint of the automotive industry \cite{ersal2020connected,Rios2018}.

The problem of safely coordinating vehicles through a merging roadway was originally addressed by the influential work of Levine and Athans \cite{Levine1966} \eat{\cite{Athans1969,Levine1966}} in which they formulated the problem as a linear quadratic regulator to minimize the speed errors that determine the distance between the merging vehicles. Following that, there have been numerous studies that tackled the problem of coordinating CAVs in different traffic scenarios, including highway merging, using techniques of classical control \cite{raravi2007merge,marinescu2012ramp,Awal2013,Ntousakis:2016aa,Rios-Torres2017b}. 
The efficacy of some of these techniques has been demonstrated through practical implementations \cite{mahbub2020sae-1,Beaver2020DemonstrationCity,chalaki2020experimental}. A thorough review of the state-of-the-art methods and challenges in coordination of CAVs is provided in \cite{Malikopoulos2016a,guanetti2018control}.

Since deriving analytical solutions for complex transportation applications is not practical, some studies used reinforcement learning techniques (RL) like Q-learning \cite{watkins1989learning} for different traffic scenarios \cite{ji2009optimal,jacob2005integrated,davarynejad2011motorway,chalaki2020hysteretic,ivanjko2015ramp,wang2019q,ngai2011multiple}. However, in large problems with many state-action pairs, to avoid Bellman's "curse of dimensionality," deep reinforcement learning methods (DRL), such as  Deep Q-network (DQN) \cite{mnih2013playing}, are used where the Q-function is replaced with a deep neural network.
There have been a few studies in the literature that have applied DRL techniques to the problem of highway merging. Wang and Chan \cite{wang2017formulation} presented a DRL formulation for on-ramp merging of an autonomous vehicle using DQN. 
The same problem of freeway merging was addressed by Nishi \textit{et al.} 
\cite{nishi2019merging} using a combination of multi-policy decision making for choosing the possible spots to merge into and passive actor-critic method to learn the state value for choosing the policy to merge into the best spot. Nassef \textit{et al.} \cite{nassef2020building} used a centralized trajectory recommendation framework for coordinating CAVs in a lane merging scenario. 
Ren \textit{et al.} \cite{ren2020cooperative} addressed the CAV merging problem in the context of a lane drop caused by a highway work zone using a soft actor-critic algorithm where only the vehicles in the merging lane were considered as RL agents, while the vehicles on the main lane were controlled by a modified VISSIM driver model. A comprehensive review of DRL methods applied to transportation research can be found in \cite{farazi2020deep}



The contribution of this paper is the development of a decentralized, multi-agent framework for safely coordinating CAVs in a highway on-ramp merging scenario while ensuring safety and smooth traffic flow.
The solution presented in this paper has the following benefits compared to other studies addressing coordination of CAVs in a merging scenario. First, this study utilizes the added benefits of connectivity by controlling a network of CAVs rather than navigating an autonomous ego-vehicle through an environment of other vehicles \cite{nishi2019merging,wang2017formulation} to safely merge into one lane. Second, we consider the CAVs as multiple cooperative learning agents as opposed to a single agent as shown in \cite{seliman2020automated}. In contrast to the application of DRL for merging of CAVs in \cite{ren2020cooperative}, this paper explicitly considers rewarding safe, high-speed travel to incentivize smoother traffic flow. Moreover, this paper considers all CAVs in the network to be RL agents and hence are required to learn to safely cooperate with other CAVs.

The layout of this paper is as follows. In Section \ref{sec:PF}, we provide our problem formulation, including the details of the adopted DRL method. In Section \ref{sec:SimSetup}, we present the simulation parameters, the type of neural networks used and how they are trained, and in Section \ref{sec:simResults}, we provide simulation results. In Section \ref{sec:con}, we draw concluding remarks and discuss potential directions for future research.

\section{PROBLEM FORMULATION}\label{sec:PF}
\begin{figure}
            \centering
            \includegraphics[width=0.99\linewidth]{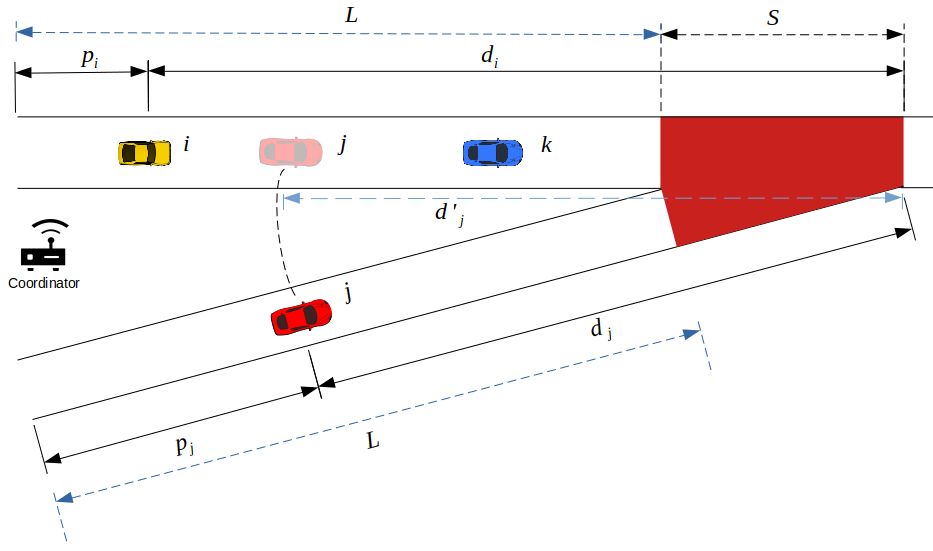}
            \caption{Merging scenario with a coordinator communicating with CAVs inside the control zone.}\label{fig:merging_schematic}%
\end{figure}

In this paper, we consider the problem of coordinating a network of CAVs in a scenario of highway on-ramp merging (Fig. \ref{fig:merging_schematic}). The problem setup consists of one group of CAVs traveling from the arterial road merging into another group of CAVs that are traveling on the main road of the highway via an on-ramp. To facilitate connectivity, in our problem, we consider the presence of a \textit{coordinator} in the network which shares information among all the CAVs without participating in any decision-making process, and also stores information specific to the environment, e.g., the physical parameters of the scenario being considered. Each CAV can share information with other CAVs and the coordinator as long as they are in a predefined area of length $L$ $\in$ $\mathbb{R}_{\plus}$ (Fig. \ref{fig:merging_schematic}) called the \textit{control zone}.  The area near the end of the control zone, where the CAVs merge, is considered as the \textit{merging zone} and its length is denoted by $S$ $\in$ $\mathbb{R}_{\plus}$ (Fig. \ref{fig:merging_schematic}). For the sake of simplicity, we assume that the highway and on-ramp are single-lane roads.

We formulate the problem as a multi-agent Markov decision process (MMDP) consisting of $N \in \mathbb{N}$ CAVs with $\mathcal{N} =\{1, 2,\cdots,N\}$ representing the set of all CAVs in the network. The problem is defined by a combination of a set of states $\mathcal{S}\colon  \mathcal{S}_{1} \times \cdots \times \mathcal{S}_{N}$ of each CAV $i$, $i \in \mathcal{N}$, in the environment; a combination of set of actions $\mathcal{U}\colon \mathcal{U}_{1} \times \cdots \times \mathcal{U}_{N}$, where $\mathcal{U}_{i}$ is the set of feasible actions to CAV $i$; and deterministic policies $\mu_{i}:\mathcal{S}_{i} \to \mathcal{U}_{i}$. The transition between states is governed by the state transition function $\mathcal{T}:\mathcal{S} \times \mathcal{U} \to \mathcal{S}$. The reward function of the entire network is denoted by $\mathcal{R}: \mathcal{S} \times \mathcal{U} \times \mathcal{S} \to \mathbb{R}$, while $\mathcal{R}_i: \mathcal{S} \times \mathcal{U}_i \times \mathcal{S} \to \mathbb{R}$ is the reward function of CAV $i$.

\subsection{Modeling Framework} 
For simulation purposes, each CAV $i\in\mathcal{N}$ is assumed to be governed by double-integrator dynamics.
        \begin{align}\label{eq:dynamics}
           \dot{p}_i(t) &= v_i(t),\\
           \dot{v}_i(t) &= u_i(t),
        \end{align}
where $p_i \in \mathcal{P}_i$, $v_i \in \mathcal{V}_i$, $u_i \in \mathcal{U}_i$ represent the position, speed, and acceleration/deceleration of CAV $i$ at time $t\in\mathbb{R}_{\plus}$. Note that our framework does not require using a specific dynamics model, and hence enables us to utilize even high-fidelity dynamics models from traffic simulators, which is the focus of ongoing work. In the context of our problem, we consider that each CAV $i$ is represented by its state $s_i$ $\coloneqq$  $\left[\bm{\mathrm{x}}_i,\bm{\mathrm{x}}_{k},\bm{\mathrm{x}}_{j}\right]^\top$, where $s_i \in \mathcal{S}_i$; $\bm{\mathrm{x}}_i\coloneqq \left [p_i,v_i\right]^\top$ consists of local information of CAV $i$ including position $p_i$ and speed $v_i$; $k$ denotes the CAV that is immediately in front of CAV $i$ on the same road as $i$, while $j$ denotes the vehicle that is ahead of $i$ on the other road. For example, in Fig. \ref{fig:merging_schematic}, if we consider the yellow vehicle to be CAV $i$, then $k$ will be the blue CAV traveling on the same road as $i$ and $j$ will be the red CAV traveling on the merging road. Additionally, let $d_{\ell}$ be the distance of any CAV $\ell \in \mathcal{N}$ from the end of the merging zone. For the CAVs traveling on the on-ramp, we project $d_{\ell}$ onto the main road, and use the projected distance to merging denoted by $d'_{\ell}$ instead. The action/control input $u_i \in [u_{\min}, u_{\max}]$ is bounded by the maximum, $u_{\max}$, and minimum, $u_{\min}$, acceleration limits of each CAV. In this study, for simplicity, we consider a homogeneous type of vehicles to represent the CAVs. Thus, $u_{\max}$ and $u_{\min}$ bounds do not vary among CAVs.

\subsection{Deep Reinforcement Learning Methodology}

Deep deterministic policy gradient (DDPG) \cite{lillicrap2015continuous} is an actor-critic RL method, which is used to learn a state-action value function (critic) and a continuous, deterministic policy function (actor), where each function is represented by a neural network. Using this architecture, consider the multi-agent scenario in which each CAV $i$ is an independent learner with its own critic and takes action independent from other CAVs. Let $s_i \in \mathcal{S}_i$ and $u_i \in \mathcal{U}_i$ represent the state and action of the CAV $i$ at the current time respectively; while $r_i \in \mathcal{R}_i$ is the reward that CAV $i$ receives at the current time for taking an action $u_i$ and transitioning from $s_i$ to $s^\prime_i \in \mathcal{S}_i$. At every time step, for each CAV $i$ the tuple $D_i = \{s_i,u_i,r_i,s^\prime_i\}$ is stored as experience buffer which is then used for training the neural networks. For each CAV, we define $Q_{i}^{\phi}(s_{i},u_{i})$ as the state-action value function and $\mu_{i}^{\theta}(s_{i})$ as the policy function, where $\phi\in[0,1]$ and $\theta\in[0,1]$ represent the corresponding neural network parameters. The loss function, $\mathcal{L}_i(\phi)$, that needs to be minimized is given by Eq. \eqref{eq:decentralized:critic} as the error between approximated state-action value function and the corresponding Bellman equation. Finally, the policy function in Eq. \eqref{eq:decentralized:critic3} is learned by maximizing the estimated optimal state-action value function. 
\begin{align}\label{eq:decentralized:critic}
&\mathcal{L}_i(\phi) = \mathop{\mathbb{E}}\limits_{s,\mu,r,s'}\left[\left(Q_{i}^{\phi}(s_{i},\mu_{i}^{\theta}(s_{i})) - y_{i}\right)^2\right],\\
\text{where},\nonumber\\
&y_{i}=r_{i} + \gamma Q_{i}^{\phi_{t}}\left(s'_{i},\mu_{i}^{\theta_{t}}(s'_{i})\right), \label{eq:decentralized:target}\\
&\mu_{i}^{\theta}(s_{i}) = \argmaxA \limits_{\theta}\mathop{\mathbb{E}}\limits_{s}\left[Q_{i}^{\phi}(s_{i},\mu_{i}^{\theta}(s_{i}))\right],\label{eq:decentralized:critic3}
\end{align}
Here $\phi_t\in[0,1]$ and $\theta_t\in[0,1]$ are target critic and target actor network parameters respectively. In DRL, target networks are used to improve stability in training. Both the target state-action value network and the target policy network are similar to their respective original networks but are updated differently which is described in Section \ref{subsec:training}.
Note that the parameters of the four neural networks are specific to each CAV and hence should also be indexed by $i$, but this has been omitted for the sake of cleaner notation.

There is one key drawback to considering all the CAVs as independent learners. As the CAVs learn their state-action value functions independently, their policies keep changing in the training procedure. This creates a non-stationary environment from the perspective of any CAV which violates Markovian assumptions that are necessary for convergence. This formulation also degrades the gradient estimates of the state-action value function that are required for maximizing the function. 
To circumvent the aforementioned issues, Lowe \textit{et al.} \cite{lowe2017multi} proposed a modification to the traditional actor-critic algorithm by considering a centralized critic and decentralized actors. The critic of each CAV is supplied with the extra information from all other CAVs to form a centralized critic. This additional information, which is only provided during the training process, includes all the actions of other CAVs and can also include the observations of other CAVs. The centralized state-action value function, using the additional information, computes the state-action value of each CAV $i$. In this formulation, since each CAV knows the actions of all other CAVs, the environment to estimate the state-value function is now stationary. The revised formulation is given by
\begin{align}\label{eq:centralized:critic}
    &\mathcal{L}_i(\phi) =\notag \\ &\mathop{\mathbb{E}}\limits_{s,\mu,r,s'}\left[\left(Q_{i}^{\phi}(s_{1},\cdots,s_{N},\mu_{1}^{\theta}(s_{1}),\cdots,\mu_{N}^{\theta}(s_{N}))  - y_{i}\right)^2\right],
\end{align}
where,
\begin{align}
y_{i}&=r_{i} + \gamma Q_{i}^{\phi_{t}}\left(s'_{1},\cdots,s'_{N},\mu_{1}^{\theta_{t}}(s'_{1}),\cdots,\mu_{N}^{\theta_{t}}(s'_{N})\right).
\end{align}
\subsection{Reward Function}
Next, we provide a detailed description of different rewards/penalties imposed on each CAV to encourage distinctive individual and cooperative behaviors. CAV $i\in\mathcal{N}$ gets a reward $r_i \in \mathcal{R}_i$ when it transitions from state $s_i\in\mathcal{S}_i$ to next state $s_i'\in\mathcal{S}_i$ by taking an action $u_i\in\mathcal{U}_i$. 
We compute the reward at the current time step by the weighted sum of individual rewards (penalties) that CAV $i$ receives to encourage (discourage) each of the following behaviors.

    	The speed of each CAV $i$ needs to be within a certain range [$v_{\min}$, $v_{\max}$]. To enforce CAV $i$ to learn and satisfy this constraint, we penalize (negative reward) any speed limit violation using $r_{i}^{\text{speed}} = -10, ~~ \text{if} ~~ v_i \geq v_{\max} ~~ \text{or} ~~ v_i \leq v_{\min}$. 
    
        We encourage increased traffic throughput by rewarding CAVs for traveling as close the maximum speed limit as possible. This is defined as
            \begin{equation}\label{highspeed}
            		r_{i}^{\text{speed}} = \frac{v_{\text{max}} - \sqrt{(v-v_{\text{max}})^2}}{v_{\text{max}}}.
        	\end{equation}
    	One critical aspect that the CAVs need to learn is to take actions that avoid collisions. For CAVs on the same road, we enforce this by penalizing rear-end collisions using $r_{i}^{\text{rear}} = - \dfrac{1}{(p_{k}-p_{i})}, ~~\text{if}~~ 0 < p_{k}-p_{i} < d_{\text{safe}},$
    where $k$ is the CAV in front of CAV $i$ on the same road, so $p_i<p_k$.
        
        To guarantee lateral safety as CAVs cross the merging zone, we impose $r_{i}^{\text{lateral}}=-\dfrac{1}{(d_{i}-d_{j})}$ for CAVs $i$ and $j$ traveling on different roads $\text{if}~~ d_{j} < S~~ \text{and}~~ d_{i} < S ~~ \text{and}~~ 0 < d_{i}-d_{j} < d_{\text{safe}}$.       		
The first two conditions verify whether CAV $i$ is in the merging zone at the same time as the vehicle in front of it, e.g., CAV $j$, while the third condition verifies whether a collision has occurred. Note that in this case, CAVs $i$ and $j$ belong to different roads and $j$ is ahead of $i$ so $d_{j}<d_{i}$. As mentioned previously, if a CAV is traveling on the on-ramp, we use its projected distance to merging, so $d_{j}$ or $d_{i}$ can be $d'_{j}$ or $d'_{i}$ respectively, depending on which one is on the merging road.

The total reward that CAV $i$ receives at the current time step is given by $r_i=w_1\cdot r_i^{\text{speed}}+w_2\cdot r_i^{\text{rear}}+w_3\cdot r_i^{\text{lateral}}$,
where $w_1,w_2,w_3 \in \mathbb{R}_{\plus}$ are the weighting factors for individual rewards. 
\section{SIMULATION SETUP}\label{sec:SimSetup}
In this section, we provide the specifics about the parameters of the different modules that are essential to execute the simulation intended to solve the problem presented in this paper. The simulation environment required for training the RL agents was custom-built in Python specifically for the highway merging scenario.

\subsection{Neural Networks} 
In the formulation adopted for this study, each CAV $i$ needs four different deep neural networks representing the critic, target-critic, actor, and target-actor networks. For all these networks, we use feedforward neural networks, and initialize the original and target networks identically. The decentralized actor network takes the state information as the input features so the number of units in its input layer is equal to the dimension of the state vector $s_{\text{dim}}=6$. Since the centralized critic needs both the state and action information of all the CAVs the number of units in its network's input layer depends both on the dimension of the action vector $u_{\text{dim}}=1$ and the number of CAVs $N$: $(s_{\text{dim}}+u_{\text{dim}}) \cdot N$. The rest of the hyperparameters that are common for both type of networks are given in Table \ref{tab:DNN_hyper}.
\begin{table}[h]
\caption{Neural Networks Hyperparameters.}
\label{tab:DNN_hyper}
\begin{center}
\begin{tabular}{|c||c||c|}
\hline
\multicolumn{1}{c}{Parameter} &\multicolumn{1}{c}{} &\multicolumn{1}{|c}{Value}\\
\hline
\hline
\multicolumn{1}{c}{Units in hidden layers 1 and 2 ($n_{h}^{1}$, $n_{h}^{2}$)} &\multicolumn{1}{c}{} &\multicolumn{1}{|c}{64}\\
\hline
\multicolumn{1}{c}{\multirow{3}{*}{Initial weights}} &\multicolumn{1}{|c}{Input layer} &\multicolumn{1}{|c}{$\mathrm{U}(-w_1,w_1)$}\\
\cline{2-3}
\multicolumn{1}{c}{} &\multicolumn{1}{|c}{Hidden layer 1} &\multicolumn{1}{|c}{$\mathrm{U}(-w_2,w_2)$}\\
\cline{2-3}
\multicolumn{1}{c}{} &\multicolumn{1}{|c}{Hidden layer 2} &\multicolumn{1}{|c}{$\mathrm{U}(-w_3,w_3)$}\\
\hline
\multicolumn{1}{c}{Optimizer learning rate ($\alpha$)} &\multicolumn{1}{c}{} &\multicolumn{1}{|c}{0.3}\\
\hline
\end{tabular}
\end{center}
\end{table}
We initialize the weights of input layer, first and second hidden layers using $w_1=w_2=1$, $w_3 = 3\times10^{-3}$ respectively, and employ Adam-PyTorch as our optimizer. We consider ReLU as the activation function for the input layer and hidden layers for both actor and critic networks. Since the output of the actor networks needs to be within the acceleration/deceleration limits, we consider the activation function of its output layer to be the $\tanh$ function, while the output layer activation function of the critic network is a simple linear function. 
\subsection{Training}
\label{subsec:training}
In order to learn the required control policies, we run the simulation for $15,000$ episodes, where the episode length is $50$ s with a time step of $0.1$ s, and if there is any collision in the current episode, we terminate it and start a new episode. At the start of every episode, CAVs are initialized close to the start of the control zone, on both the main and the merging roads, with speeds uniformly sampled from $[v_{\text{min}}, v_{\text{max}}]$. If all the CAVs in the network cross the merging zone without any collisions, we give an additional reward to each CAV in the network. During training, the neural networks of each CAV are updated periodically based on samples of experience drawn uniformly at random from the buffer of stored samples. After updating the original networks, we employ Polyak-Ruppert averaging \cite{polyak1990new} to update the target networks using a tunable polyak parameter $\tau=0.01$, which causes a gradual update in the target networks to improve stability in training.

\subsection{Simulation Parameters}
We use the following simulation parameters: $v_{\min}$ = $5$ m/s, $v_{\max}$ = $15$ m/s, $u_{\min}$ = $-3$ m/s$^2$, $u_{\max}$ = $3$ m/s$^2$, $L$ = $90$ m,  $S$ = $10$ m, $d_{\text{safe}}$ = $0.5$ m. The speed limits are chosen based on the length of the control zone and the episode length such that meaningful interactions between the agents can occur before they exit the merging zone. The acceleration limits are typical for a common passenger vehicle. The weighting factors in the reward function are set as $w_1=1$, $w_2=w_3=20$ to give a higher priority to avoid collisions when compared to the reward gained from going close to $v_{\max}$.

\section{SIMULATION RESULTS}\label{sec:simResults}
\begin{figure}[tb]
    \centering
    \includegraphics[width=\linewidth]{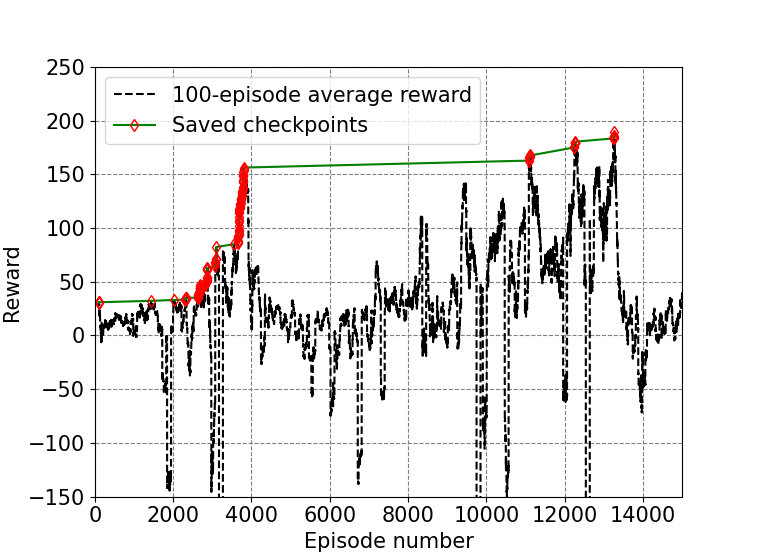}
    
    \caption{100-episode average rewards for the entire network.}
    \label{fig:avg_rew}
\end{figure}
After the training phase, we examine the learning performance by considering the 100-episode average rewards collected by all the CAVs, and this is shown in Fig. \ref{fig:avg_rew}. The rewards collected by the CAVs fluctuate during the training because there is no explicit exploitation strategy involved. This behavior is common in actor-critic methods; however, as it can be seen in Fig. \ref{fig:avg_rew}, there is an upward trend in 100-episode average rewards showing the improvement in learning. To keep a record of the policies that resulted in accumulating higher rewards, we save the actor network parameters whenever the current 100-episode average reward is better than the previous best 100-episode average reward.

In following, we use three scenarios to demonstrate the learned behaviors. 
The policies learned from our framework are only responsible for controlling the CAVs while they are in the control zone and merging zone. After leaving the merging zone, the CAVs cruise at the same speed that they exited the merging zone \footnote{Videos from our simulation analysis can be found at the supplemental site, \url{https://sites.google.com/view/ud-ids-lab/MADRL}.}.

\begin{figure}[ht]
\centering
    \begin{subfigure}{0.5\linewidth}
    \centering
    \includegraphics[width=\linewidth]{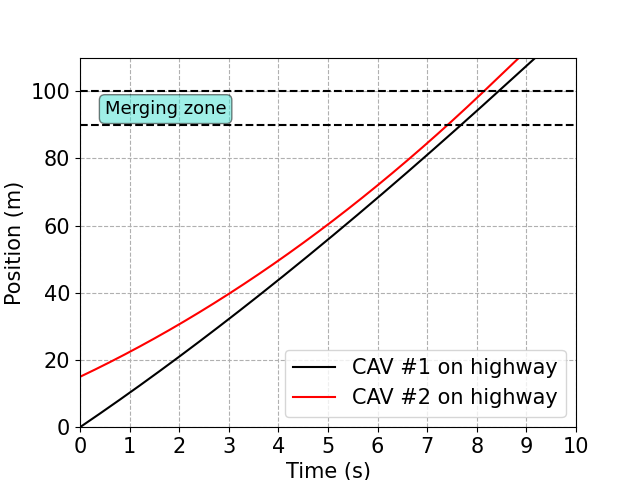}
    \caption{Position}
    \label{fig:rear_pos}
    \end{subfigure}%
    \begin{subfigure}{0.5\linewidth}
    \centering
    \includegraphics[width=\linewidth]{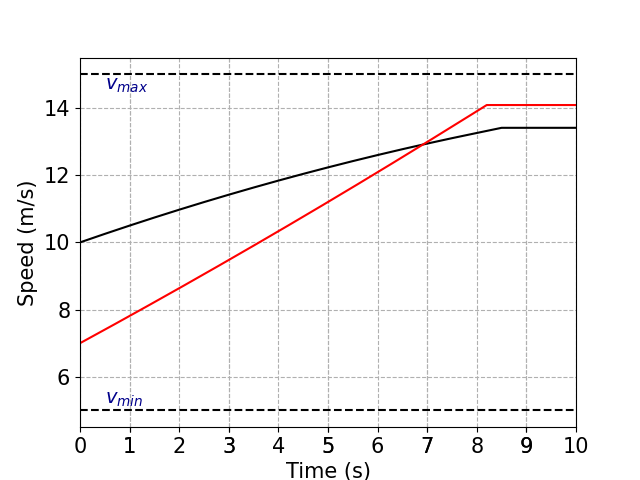}
    \caption{Speed}
    \label{fig:rear_vel}
    \end{subfigure}\\[1ex]
    \caption{Position and speed trajectories for rear-end collision avoidance scenario.}
    \label{fig:rear}
    \end{figure}

To demonstrate rear-end collision avoidance behavior learned during the training process, we initialize two CAVs on the highway such that CAV \#$1$ which is behind CAV \#$2$ has a higher initial speed. The position and speed trajectories of two CAVs are shown in Fig. \ref{fig:rear}. As it can be seen in Fig. \ref{fig:rear_vel}, both CAVs are incentivized to increase their speed to $v_{\text{max}}$. However, due to the rear-end safety constraint, CAV \#$1$ increases its speed at a lower rate to avoid the collision.

\begin{figure}[ht]
    \centering
    \begin{subfigure}{0.5\linewidth}
    \centering
    \includegraphics[width=\linewidth]{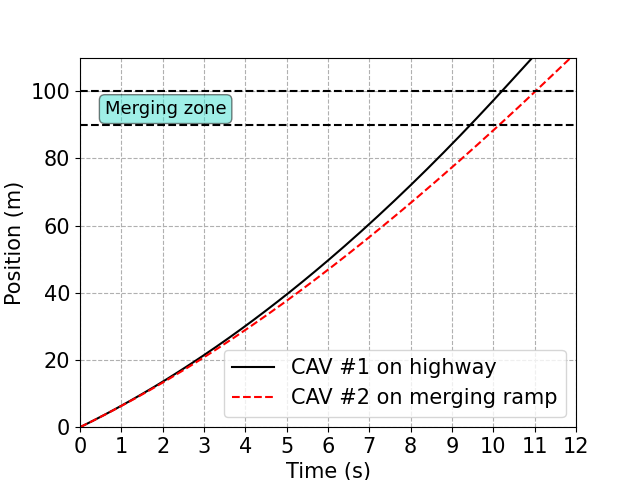}
    \caption{Position}
    \label{fig:lat_pos}
    \end{subfigure}%
    \begin{subfigure}{0.5\linewidth}
    \centering
    \includegraphics[width=\linewidth]{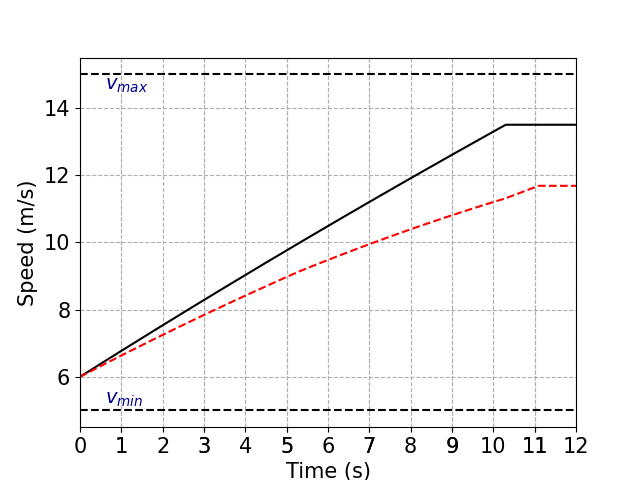}
    \caption{Speed}
    \label{fig:lat_vel}
    \end{subfigure}\\[1ex]
    \caption{Position and speed trajectories for lateral collision avoidance scenario.}
    \label{fig:lateral}
    \end{figure}

For the lateral collision scenario, we initialize one CAV on the highway and one on the on-ramp with the same initial speeds. The position and speed trajectories of two CAVs are shown in Fig. \ref{fig:lateral}. As it can be seen in Fig. \ref{fig:lat_vel}, CAV \#$2$ on the merging road slows down in order to avoid lateral collision in the merging zone. Additionally, as CAV \#$1$ exits the merging zone around $10$ s, CAV \#$2$ increases its speed with a higher rate. 

\begin{figure}[ht]
    \centering
    \includegraphics[width=\linewidth]{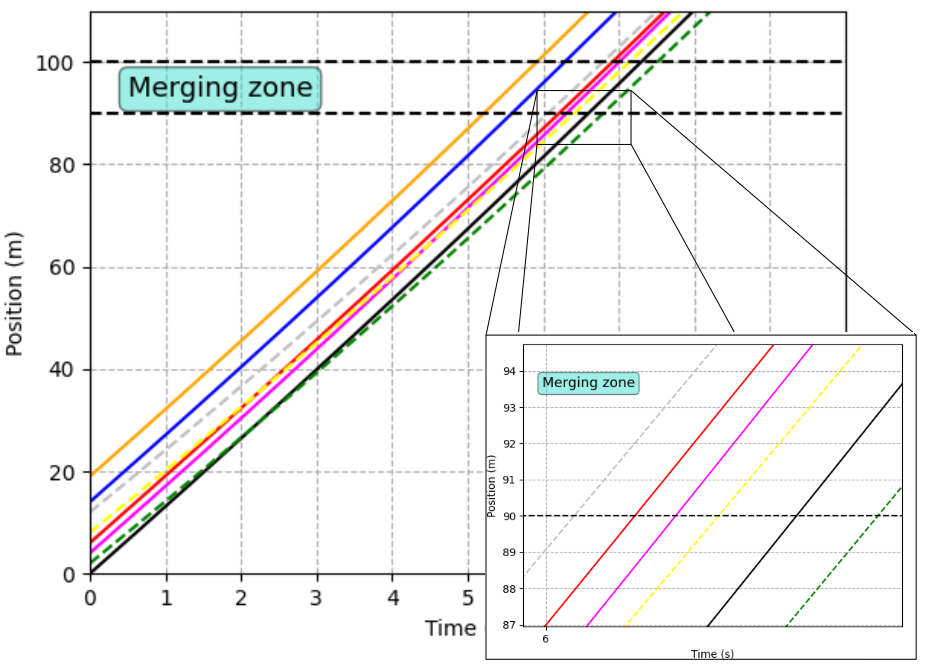}
    
    \caption{Position trajectories of 8 CAVs in the network.}
    \label{fig:rear_and_lateral}
\end{figure}

For the final scenario, to demonstrate the satisfaction of rear-end safety and lateral safety constraints when there are more CAVs present in the network, we consider coordination of $8$ CAVs.
We pick the initial position of these CAVs randomly, while the initial speed of CAVs on the highway and on-ramp roads are set to $13$ m/s and $12$ m/s, respectively.  
Fig. \ref{fig:rear_and_lateral} illustrates the position trajectories of the CAVs on the highway (solid lines) and on the on-ramp road (dashed lines). As it can be seen from the figure, the CAVs learned to satisfy the rear-end safety constraint from the initial time until the time they exit the merging zone. Additionally, trajectories of CAVs from different paths do not intersect inside the merging zone, showing the satisfaction of the lateral safety constraint. 
The close-up view in Fig. \ref{fig:rear_and_lateral} presents the position trajectories of the CAVs inside the merging zone, confirming that the CAVs indeed avoid lateral collisions in the merging zone by maintaining distance more than $d_{\text{safe}}$ with their respective front vehicles. 

For this scenario, we initialize the actor network parameters of each CAV with trained network parameters of one CAV from a total of three CAVs involved in the training process.
This shows that we can use our framework to train the system for some finite number of vehicles, and transfer the learned policy to any number of vehicles.

\begin{figure}[tb]
     \centering
      \includegraphics[width=\linewidth]{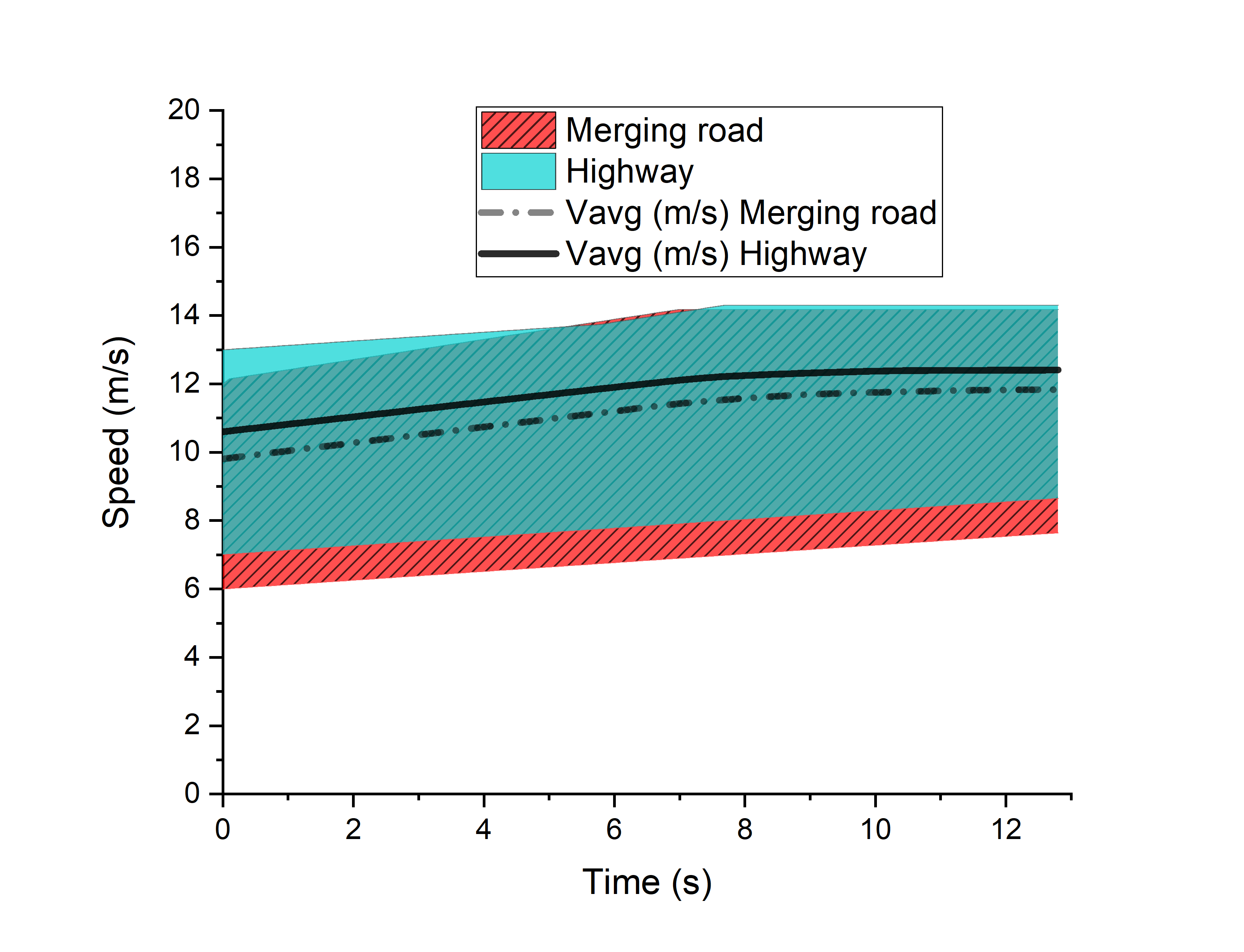}
      \caption{Speed range for merging and highway roads over 5 experiments.}\label{fig:range_avg}
\end{figure}

We further demonstrate the effectiveness of our approach in eliminating the stop-and-go driving by performing five different simulations with random initial speeds ranging from $6$ m/s to $13$ m/s. The instantaneous average, maximum, and minimum speed of CAVs inside the control zone for the merging road and highway for all five simulations are shown in Fig. \ref{fig:range_avg}. The minimum speed of CAVs for both roads over five experiments is positive indicating smooth traffic flow.

    The above scenarios clearly illustrate the efficacy of the policies learned using the reinforcement learning framework presented in this paper. The actions derived from these policies ensure improved traffic throughput as a consequence of high-speed travel, but simultaneously adhering to the set speed limits, along with safe coordination to execute merging maneuver without any rear-end and lateral collisions.
    
\section{CONCLUDING REMARKS AND DISCUSSION}\label{sec:con}
In this paper, we proposed a decentralized, multi-agent reinforcement learning-based framework for coordinating CAVs through a highway merging scenario. In our framework, we employed an actor-critic architecture with a centralized critic and decentralized actors to avoid the problem of a non-stationary environment \cite{lowe2017multi} induced by decentralized learning CAVs. Our framework enables the capability of implementing the learned policies in any number of CAVs by transferring the policies which were learned through a few learning CAVs in the training process. In addition to ensuring rear-end and lateral safety, our choice of reward function encouraged the CAVs to learn to travel at high speeds which in turn results in smoother traffic flow. Finally, we showed the effectiveness of the proposed approach through several simulations. 
As part of ongoing research, we are extending the current framework to other traffic scenarios, including urban intersections and roundabouts, while simultaneously utilizing high-fidelity dynamics models. In addition to that, we are investigating the effects of noise originating from vehicle-level control, and also of errors and delays in communication since in its current form the CAVs in our framework rely on accurate information from other CAVs, and hence are not robust to uncertainties. Our framework could also be extended to study the interaction of human-driven vehicles and CAVs in mixed-traffic scenarios.

\bibliographystyle{IEEEtran}
\bibliography{references/IDS_Publications_03202022.bib,references/ref.bib,IEEEabrv}

\end{document}